# Photoinduced giant modulation of terahertz nonlinearity from metasurfaces

Chen Wang[a, 1], Yong Tan[a, 1], Yongzheng Wen[a, ]*, Shiqiang Zhao[a], Kaixin Yu[a], Renfei Zhang[b], Jingbo Sun[a], and Ji Zhou[a, ]*

[a] State Key Laboratory of New Ceramic Materials, School of Materials Science and Engineering, Tsinghua University, Beijing 100084, People's Republic of China.

[b] Research Center for Metamaterials, Wuzhen Laboratory, Jiaxing 314500, People's Republic of China.

*Corresponding authors at: State Key Laboratory of New Ceramic Materials, School of Materials Science and Engineering, Tsinghua University, Beijing 100084, People's Republic of China (Y. Wen and J. Zhou)

*E-mail addresses*: Wen, Y. (wenyzheng@tsinghua.edu.cn), Zhou, J. (zhouji@tsinghua.edu.cn).

[1] These authors contributed equally to this work.

**Abstract**

Active control of optical nonlinearity is essential for advancing next-generation electronics and photonics, including high-speed wireless communications, optical information processing, and nonlinear signal manipulation. However, achieving tunable nonlinearity at terahertz (THz) frequencies faces significant challenges due to the lack of materials that combine high nonlinear responses with strong sensitivity to external stimuli in this spectral regime. Here, we show giant modulation of THz nonlinearity by optically tailoring the valley degree of freedom in semiconductor-based metasurfaces. Mediated by the resonant behaviors of metasurfaces, photoexcited electrons transition into different valleys in the conduction band in response to the driving THz field, with the transition rate controlled by light intensity. Since THz nonlinearities vary significantly with electron dynamics in different valleys, various nonlinear effects—such as nonlinear transmission and generation—can be efficiently enhanced and modulated within a single metasurface using weak optical pumping. With optical energy as low as a few picojoules, we achieve on-off switching of THz third harmonic generation with a modulation depth exceeding $2\times10^4$%, along with effective tunability of its nonperturbative behaviors. Our approach breaks new ground in active THz devices fully compatible with semiconductor industry standards, indicating a promising building block for ultrafast THz signal processing, all-optical computing, and nonlinear optical elements.

# 1 Introduction

Dynamic control of the optical nonlinearity of a material provides exclusive freedoms to manipulate light and lays the foundations of myriad intriguing photonics technologies, including optical switching[1,2], reconfigurable frequency multipliers[3], compact quantum sources[4,5], and optical neural networks[6,7]. Ideal materials for controlling optical nonlinearity should meet several key criteria: they should exhibit large nonlinear responses, offer high modulation depth under external stimuli, and enable multidimensional modulation. Other properties including low power consumption, ease of implementation, room-temperature operation, and compatibility with integrated platforms are also highly desired in practice. Progresses have been made in pursuing one or more of these ambitious objectives within the visible, infrared, and microwave spectral regimes, where abundant nonlinear materials and devices are available[8-11]. Considerable efforts are still ongoing to discover materials that can simultaneously fulfill all these requirements. However, the quest for such control at the terahertz (THz) frequencies remains challenging and hardly explored, because of the scarcity of the materials exhibiting efficient THz nonlinearity alongside external controllability. Given the importance of THz frequencies as the unique band bridging electronics and photonics, and its recognized potential in fields like next-generation communications[12,13], radar[14], and medical diagnosis[15,16], addressing this limitation has become increasingly imperative for unlocking the full capabilities of THz technology.

Recent efforts have been devoted to investigating the THz nonlinearity in advanced materials such as Fermionic Dirac matters[17] and superconductors[18], but these materials largely suffer from inefficient nonlinear responses or limited tunability. For instance, graphene has preliminarily demonstrated the modulation of THz third harmonic generation (THG) via tuning the Fermi level with electrical gating[19], but its atomic-level thickness leads to low efficiency and limited modulation depth, and the electrical tunability suffers from the bulky power supplies and complex control circuits. Optical control, in contrast, offers superior advantages like circuit-free design, non-contact operation, high scalability, and ultrafast responses[20-23]. While pioneering work has explored the optically controlled THz nonlinearity in superconductors based on the order-parameter change and quasiparticle excitations, its modulation depth is very limited even at cryogenic temperatures[24]. Furthermore, both graphene and superconductors face difficulties in modulating multiple nonlinear effects and integration into compact systems.

As excellent platforms for building compact THz devices, semiconductors, like silicon (Si) and germanium, have been theoretically predicted to possess large THz nonlinearity due to their unique advantages in multivalley interactions and small effective mass[25,26]. While such theoretical predictions have been partially validated in measurements carried out with complex setup involving cryostat and free-electron lasers[27,28], the experimental demonstrations at room temperature are notably absent, let alone their active modulation. Metasurfaces, known for their ability to confine light at subwavelength scales and enhance light-matter interactions, offer a potential solution to amplify THz nonlinear responses[29-31]. Despite this potential, effective methods for actively modulating these responses remain elusive. Theoretical models suggest that the transport dynamics of free carriers, which can be substantially tailored by the valley degree of freedom, govern the THz nonlinear responses[32]. Carrier populations in different valleys exhibit distinct properties due to the electronic band structures, resulting in different transport behavior and therefore different nonlinear responses. However, there is no report on realizing valley-specific THz nonlinearity, nor any demonstration of modulating the nonlinearity by controlling the intervalley transitions.

In this work, we propose a strategy for achieving giant modulation of THz nonlinearity in multivalley semiconductor-based metasurfaces. Mediated by the resonant enhancement of metasurfaces, we can simultaneously engineer the key factors governing THz nonlinearity through optical pumping, especially the excitation of free carriers and their population in different energy valleys. By precisely controlling over the valley degree of freedom, this approach enables substantial switching and modulation of various THz nonlinear effects, including nonlinear transmission and THG, within a single metasurface using only weak optical intensity. We experimentally demonstrate on-off switching and modulation of THz THG with a modulation depth exceeding $2\times10^4$% under the optical energy inputs as low as a few picojoules. Furthermore, we achieve control of nonperturbative behaviors of nonlinearity, which was rarely explored but holds critical potential for all-optical computing and THz signal processing. Our work paves the way for developing efficient THz nonlinear devices with large tunability, and full compatibility with semiconductor industry standards.

## 2 Results

### 2.1 Principle of the giant modulation of THz nonlinearity

The designed semiconductor-based metasurface is illustrated in Fig.1a. The THz nonlinearities arise from the interactions of the free electrons in semiconductors with the driving THz field, which can be described by the third-order nonlinear polarization density, $P_{NL}^{(3)}$, as

$$P_{NL}^{(3)} = \varepsilon_0(N_X\alpha_X^{(3)} + N_L\alpha_L^{(3)})E_{THz}^3, \tag{1}$$

where $N_X$ ($N_L$) and $\alpha_X^{(3)}$ ($\alpha_L^{(3)}$) are the free electron density and the third-order nonlinear polarizability of a single electron at $X$ ($L$) valley of conduction band, respectively, $E_{THz}$ is the local THz field, and $\varepsilon_0$ is the vacuum permittivity. This equation highlights that the presence of free carriers, the population of carriers at different energy valleys, and the THz field strength are critical factors in determining the THz nonlinear responses, all of which can be effectively engineered by optical pumping, mediated by the resonant behavior of metasurfaces. Considering a multivalley semiconductor such as Si, light with photon energy exceeding the bandgap excites bound electrons from the valence band into the X-valley of the conduction band. Metasurfaces provide the local enhancements on the THz electric field, leading to strong nonlinear responses rooting from these photoexcited free electrons, which can be optically switched. Meanwhile, driven by the enhanced fields, the X-valley free electrons gain sufficiently large ponderomotive energy to overcome the intervalley energy barrier, resulting in pronounced transition into the L-valley. The nonlinear transports of the electrons at different valleys significantly differ due to the distinct nonparabolicity of the band[33,34], leading to the variations of nonlinear effects. Given a certain amplitude of the THz pump, the intervalley transition rate, that is the population of free electrons at different valleys, can be finely controlled by tuning the optical pump intensity, as it influences the resonance and the field enhancements of the metasurface by altering the density of the newly-generated electrons and the dielectric properties of Si. Therefore, by simultaneously switching and engineering all essences of THz nonlinearity, even weak light intensity can lever giant modulation of various THz nonlinear properties in a single semiconductor-based metasurface.

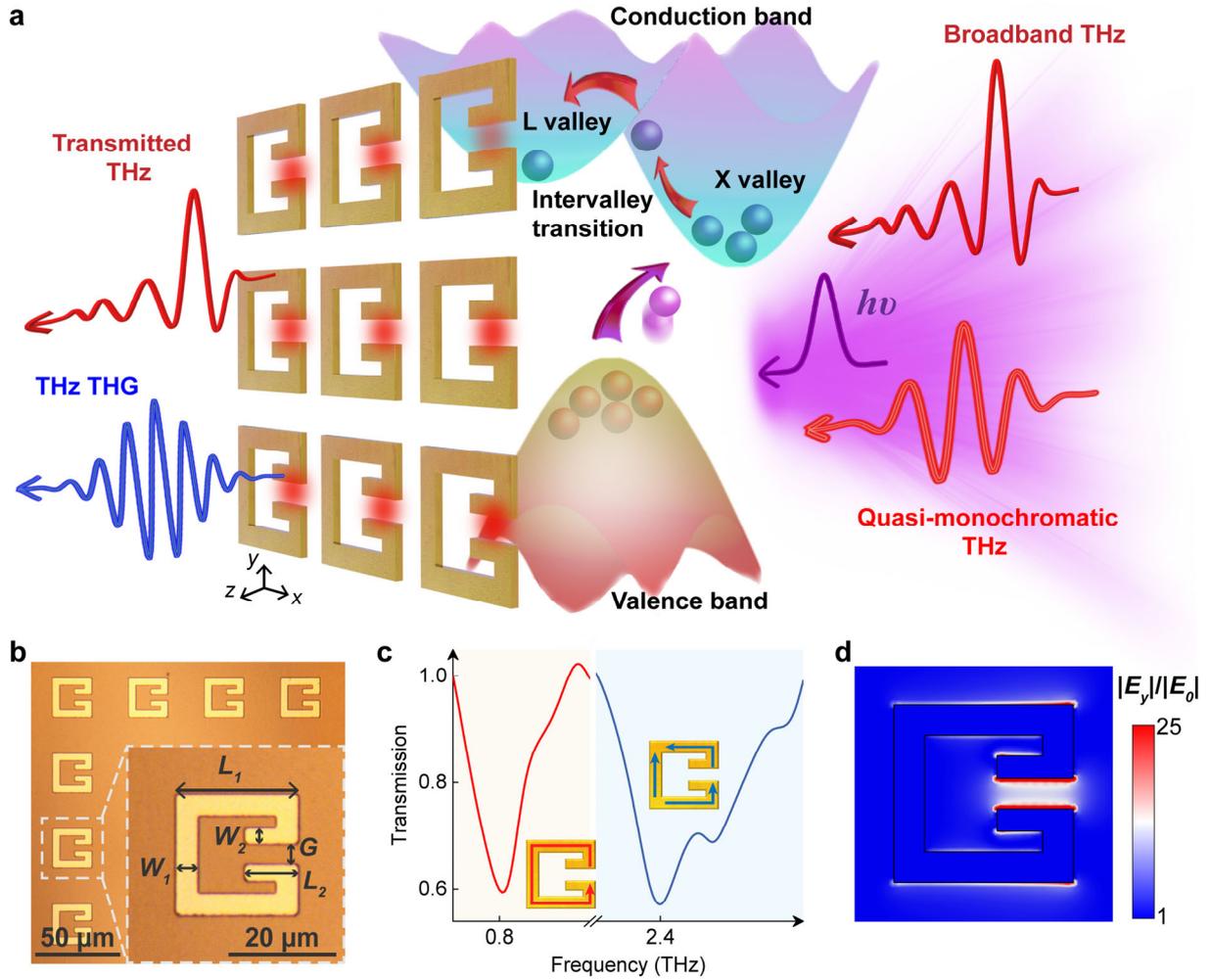

**Fig.1. Photoinduced giant modulation of THz nonlinearities in semiconductor-based metasurface.** (**a**) The physical principle of the photoinduced modulation of THz nonlinearities. (**b**) Microscopic image of the metasurface with a lattice constant of 45 μm, where the inset shows a single unit cell: $L_1$=23 μm, $L_2$=10 μm, $W_1$=4 μm, $W_2$=3 μm, $G$=4 μm. (**c**) Measured transmission spectra of the metasurface without optical pump beam. The insets show the surface current of the resonances. (**d**) Simulated distribution of the y-component electric field, $|E_y|$, of the metasurface ($\sigma_{Si}$=1 S m$^{-1}$) at 0.8 THz, normalized to the amplitude of the incident, $|E_0|$.

Guided by the theory, we designed and fabricated a proof-of-principle metasurface composed of a periodic arrangement of aluminum split-ring resonators (SRRs) on a single crystalline undoped Si film (Fig. 1b). The SRRs exhibit resonances at 0.8 THz and 2.4 THz, where the lower-order resonance enhances the THz field at the fundamental frequency, and the higher-order resonance facilitates efficient generation of THz third harmonic[35,36] (Fig. 1c). Near-infrared light with photon energy of *hν*=1.55 eV was selected to pump the metasurface, enabling generation of carrier in Si. We considered two THz waveforms, single-cycle broadband and multicycle quasi-monochromatic THz pulses, to demonstrate active modulation of two typical nonlinear phenomena: THz field-strength-dependent transmission and THG. The finite element method was employed to model and simulate the metasurface, with the local THz electric field distribution at 0.8 THz depicted in Fig. 1d (see Note S1 in Supplementary Material for simulation settings). Due to the capacitive effect formed by the gaps in SRRs, significant enhancements in the THz electric field were achieved.

## 2.2 Modulation of THz nonlinear transmission

Under the local intense THz field, the metasurface exhibits nonlinear transmission. The strong dependence of this nonlinear transmission on the conductivity of the metasurface allows for straightforward and efficient modulation through optical pumping. Building on the theoretical framework, we quantitatively investigate the THz nonlinear transmission of the metasurface and its photoinduced modulation through experimental measurements. We first studied the THz field-strength-dependent transmission in the metasurface. As shown in Fig. 2a, after conduction electrons are accelerated by the THz field, part of the ponderomotive energy gained is released during inelastic collisions with valence electrons, promoting them to the conduction band[37,38]. This cascade effect, which results in an avalanche of free electrons, is known as impact ionization. The carriers generated through impact ionization significantly modify the THz transmission in the metasurface. We characterized this process in the sample using a THz time-domain spectrometer (see Methods and Note S2 in Supplementary Material for details). Figure 2b shows the metasurface transmission under various single-cycle THz fields without the presence of near-infrared pump light. At the lowest field, impact ionization is not significant, and a characteristic dip corresponding to the resonance of the SRRs is observed. As the incident field increases, the resonance frequency redshifts, and the transmission increases and broadens. The transmission of the metasurface under THz field strength of 68 kV cm$^{-1}$, without optical pumping, is denoted as $T_0$. To quantitatively assess the effect of THz field strength on metasurface transmission, we introduced the modulation depth of transmission ($MD_T$), defined as $MD_T = (T - T_0)/T_0 \times 100\%$, where $T$ represents the transmission of the metasurface. As shown in Fig. 2c, the resonance of the metasurface is highly sensitive to changes in the conductivity of the Si, resulting in a modulation depth of up to 171% at the resonant frequency of 0.84 THz. To elucidate the nature of nonlinear transmission modulation, we simulated the metasurface by modeling Si with a THz field-strength-dependent conductivity. As shown in Fig. 2d, the simulation results exhibit excellent agreement with the experimental measurements, confirming that the observed nonlinear transmission behavior arises from the stimulated carrier, whose density is positively correlated with the THz field strength[37,39].

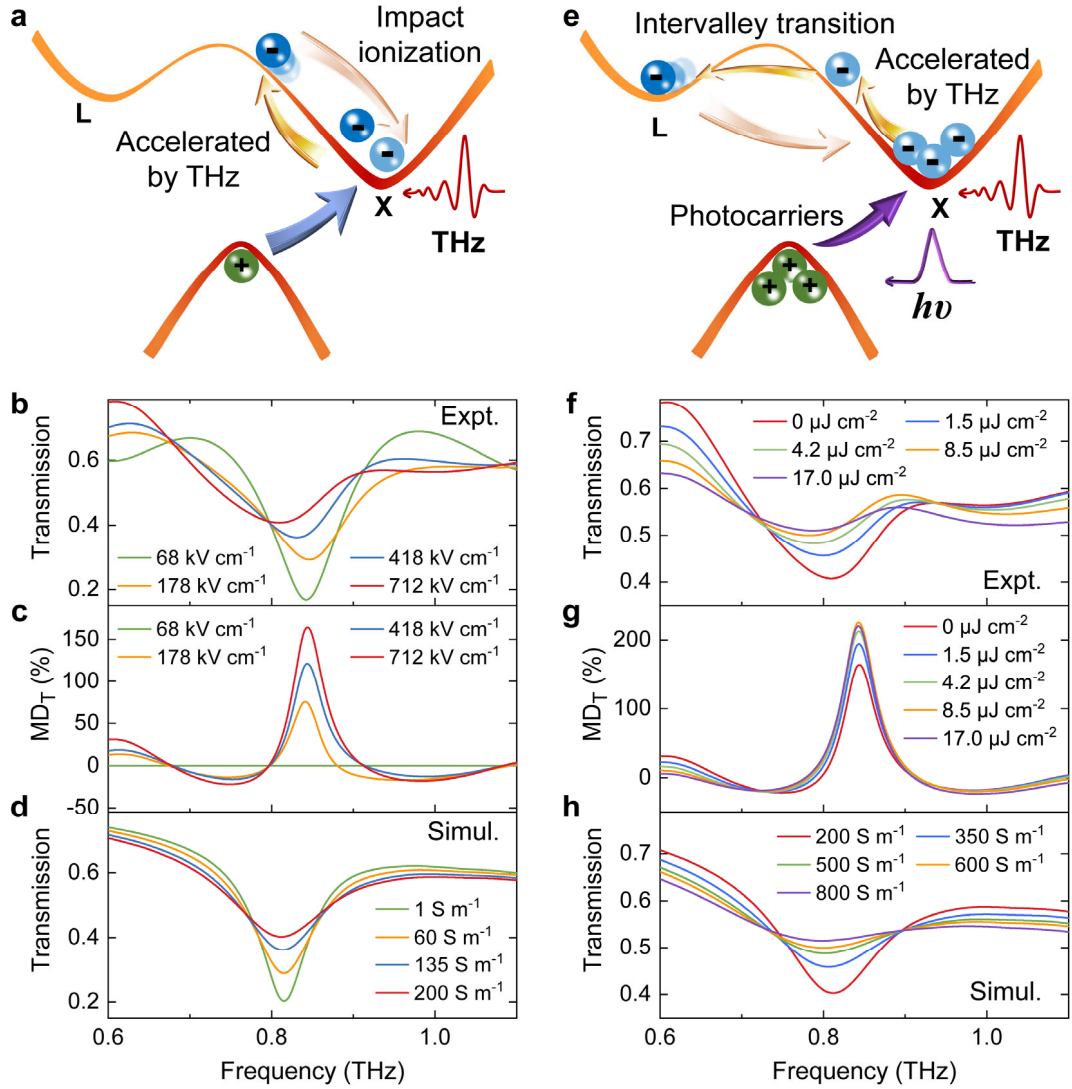

**Fig. 2. Photoinduced modulation of THz nonlinear transmission.** (**a**) Schematic of THz-induced impact ionization in Si. Measured nonlinear transmission spectra through the metasurface (**b**) and modulation depth (**c**) for various incident single-cycle THz peak electric fields. (**d**) Simulated transmission spectra with various Si conductivities. (**e**) Schematic of photogenerated carriers and THz-induced intervalley transition. Measured nonlinear transmission spectra (**f**) and modulation depth (**g**) of the metasurface under various optical pump fluences with the THz peak electric field as 712 kV cm$^{-1}$. (**h**) Simulated transmission spectra with various Si conductivities.

Such nonlinear transmission, closely related to the conductivity of the metasurface, can be further modulated using photonic methods. As shown in Fig. 2e, the large number of carriers generated by the near-infrared light shortens the kinetic energy acceleration process, which inhibits the avalanche multiplication effect, thereby suppressing impact ionization. At this stage, the photogenerated carriers acquire sufficient energy from the THz field to overcome the potential barrier between energy valleys and scatter to higher-energy satellite valleys, leading to intervalley transition[32,39,40]. At a peak field strength of 712 kV cm$^{-1}$, intervalley transition causes the photogenerated carrier density in the L-valley to be approximately 10 times higher than in the X-valley. This redistribution of carriers between energy valleys significantly reduces the conductivity of Si compared to the case without intervalley transition (see Note S3 in Supplementary Material for details). We measured the nonlinear transmission as a function of pump fluence, with key experimental results presented in Fig. 2f. The peak field strength of the THz pulse was maintained at 712 kV cm$^{-1}$. The generation of numerous photogenerated carriers makes the gap of the metasurface conductive, resulting in a shallower resonant transmission dip. As shown in Fig. 2g, at a pump fluence absorbed by the metasurface of 8.5 μJ cm$^{-2}$ (see Note S4 in Supplementary Material for details on metasurface absorption), we achieved a maximum modulation depth of 237% at the resonance. The simulated spectra, shown in Fig. 2h, which account for the variation in Si conductivity due to intervalley transition, are in good agreement with our experimental results (see Note S3 in Supplementary Material for details). The slight discrepancies in resonant frequency and bandwidth between the experimental and simulated results primarily stem from fabrication imperfections. The modulation depth of our metasurface exceeds that of a recently reported high-performance modulator by more than threefold[21]. This highlights the efficient and straightforward control of the nonlinear transmission in the metasurface using low pump fluence.

**2.3 Switching and modulation of THz THG**

To verify the THz THG in the designed metasurface under optical excitation, the metasurface was excited collinearly by a quasi-monochromatic multicycle THz field with a central frequency of 0.8 THz and the same optical pulse. The peak field strength of the fundamental THz pulse was as low as 12.1 kV cm$^{-1}$ (see Note S2 and Note S5 in Supplementary Material for details). As shown in Fig. 3a, the metasurface initially exhibits an extremely weak nonlinear response due to the absence of carriers, leaving almost no THz THG signal observed (blue curve). Under the excitation of the optical pump pulse, a prominent third harmonic signal at 2.4 THz appears in the transmission spectrum (red curve). The optical pump pulse with a fluence of 4.2 μJ cm$^{-2}$ generated sufficient photogenerated carriers, and the local THz electric field enhanced the nonlinear interaction between the THz field and the metasurface. Consequently, the metasurface was switched on and generated significant third harmonics under low-threshold excitation, achieving a field conversion efficiency of $1.1\times10^{-5}$ (kV cm$^{-1}$)$^{-2}$, which is an order of magnitude higher than that of graphene[41]. To confirm that the observed THG under optical excitation arises from the physical process designed with theoretical guidance, we characterized the plain silicon-on-sapphire (SoS) substrate under identical experimental conditions. Despite the presence of sufficient photogenerated carriers in the SoS substrate, the nonlinear response of the Si was too weak for the moderate THz fundamental waves to excite THG (green curve). We also simulated the THG of the metasurface, and the results are in good agreement with the experimental data (see Note S6 in Supplementary Material for details). These findings suggest the existence of photoinduced switchable THz THG within the metasurface.

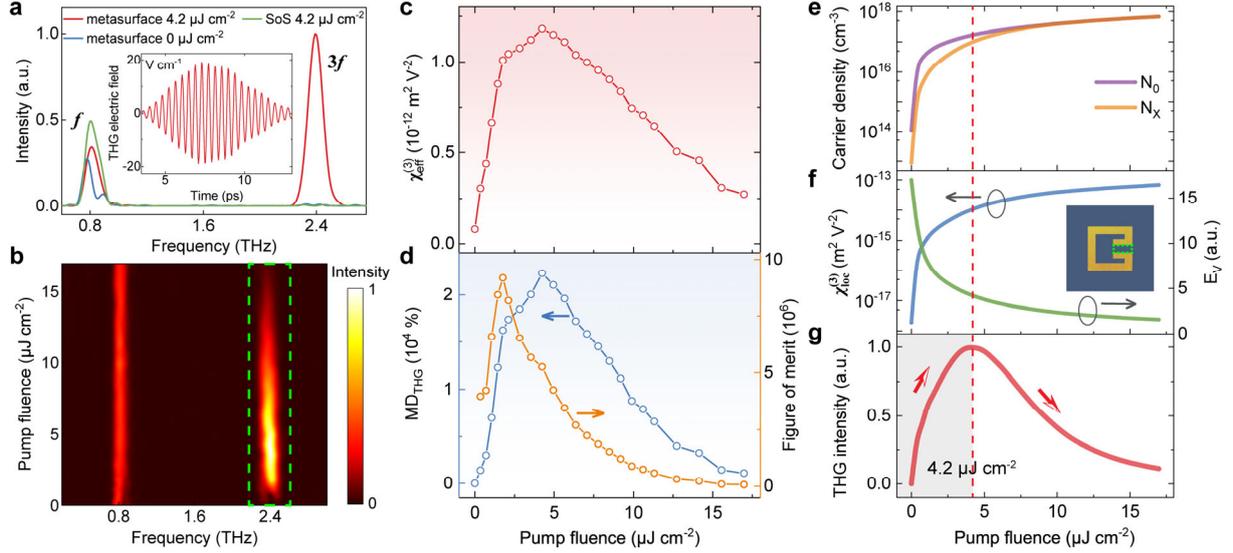

**Fig. 3. Photoinduced switching and modulation of THz THG. (a)** The measured frequency-domain spectra of the metasurface and SoS substrate at 0.8 THz fundamental pump with the 12.1 kV cm$^{-1}$ field strengths, and the optical fluences are 0 and 4.2 μJ cm$^{-2}$. The inset shows the time-domain THG spectrum of the metasurface at 4.2 μJ cm$^{-2}$ optical fluences. Measured normalized THz THG intensity (**b**), $\chi^{(3)}_{\text{eff}}$ (**c**), and $MD_{THG}$ and figure of merit (**d**) of the metasurface under various optical pump fluences. Calculated carrier density (**e**), $\chi^{(3)}_{\text{loc}}$ and $E_v$ (**f**), as well as normalized THz THG intensity (**g**) as functions of the optical pump fluences.

For our metasurface, increasing the free carrier density in Si enhances its THz nonlinearity but simultaneously weakens the local THz field strength. This indicates that the THz nonlinearity of the metasurface can be externally controlled, with an optimal photoinduced carrier density existing that maximizes its nonlinear response. As shown in Fig. 3b, the frequency-domain spectra of the THz pulses transmitted through the metasurface are presented at various optical pump fluences with a fixed THz peak field strength of 12.1 kV cm$^{-1}$. As the pump fluence increases, the intensity of the THz third harmonic initially rises, reaching a maximum at 4.2 μJ cm$^{-2}$, before decreasing. We extracted the effective third-order nonlinear susceptibility ($\chi_{\text{eff}}^{(3)}$) (see Note S7 in Supplementary Material). As shown in Fig. 3c, the maximum $\chi_{\text{eff}}^{(3)}$ in the metasurface reaches up to $1.2\times10^{-12}$ m$^2$ V$^{-2}$, which is one to two orders of magnitude larger than that of plain Si (approximately $10^{-14}$ to $10^{-13}$ m$^2$ V$^{-2}$)[31,33]. To quantify the photoinduced modulation of the THz third nonlinearity, we introduced the modulation depth of the third harmonic intensity ($MD_{TH}$), defined as $MD_{TH}=\left(I_{3f,on}-I_{3f,off}\right)/I_{3f,off}$, where $I_{3f,on}$ and $I_{3f,off}$ represent the THz third harmonic intensities with and without optical pumping, respectively. As shown in Fig. 3d, our measurements demonstrate a peak $MD_{TH}$ of up to $2.23\times10^4$% at a low pump fluence of 4.2 μJ cm$^{-2}$. To our knowledge, this represents the largest modulation depth of THz third harmonic reported to date[19,24]. To compare the performance with other systems, we also present the figure of merit (FoM) for the metasurface, defined as FoM = $MD_{TH}$ (%) / pump fluence (mJ cm$^{-2}$)[42,43]. The metasurface exhibits a giant FoM of approximately $9\times10^6$, far surpassing the value observed in three-dimensional topological insulators (~$5\times10^4$)[42]. Considering the SRR area of 23×23 μm$^2$, we estimated a modulated pump energy of 9.4 pJ for the THz THG in the unit cell of the metasurface at the peak FoM. Our metasurface features an ultrahigh modulation depth of THz THG with low modulated energy, offering a new approach for high-performance frequency multipliers and modulators in the THz frequency range.

  We theoretically elaborated on the effects of pump fluence on the metasurface, with carrier density, the population of carrier at different energy valleys, and local THz field strength acting as control parameters for tuning the nonlinear response. As shown in Fig. 3e, we first calculated the total carrier density ($N_0$) where the optical pump injects electrons from the valence band into the conduction band through interband transitions. Then, the population of carrier at different energy valleys was calculated based on the THz field-driven dynamics of the carriers (see Note S3 in Supplementary Material). As the pump fluence increases, the resonance strength and local THz electric field decrease, resulting in fewer carrier transitions, and the carrier density in the X-valley ($N_X$) approaches the total carrier density. Compared to the L-valley, the effective mass of carriers in the X-valley is lower, resulting in a greater nonlinear contribution to Si ($\alpha_X^{(3)} > \alpha_L^{(3)}$). Significant intervalley transition in the absence of optical pumping causes most carriers to transfer to the L-valley, weakening the nonlinearity compared to conditions without intervalley transition, thereby facilitating a higher third-harmonic extinction ratio and modulation regime. Since the local third-order nonlinear susceptibility ($\chi_{\text{loc}}^{(3)}$) of Si in the metasurface at THz frequency range is proportional to the carrier density (see Note S8 in Supplementary Material for details)[33], it increases with higher pump fluence, as shown in the blue line in Fig. 3f. To quantitatively investigate the effect of local THz field strength on nonlinearity, we simulated the metasurface and obtained the average enhancement factor of the local electric fields at the gap ($E_V$), as shown by the green line in Fig. 3f (see Note S1 in Supplementary Material for details). The $E_V$ decreases with increasing pump fluence. The third harmonic intensity generated by the metasurface follows the relation: $I_{3f} \propto \left(\chi_{\text{loc}}^{(3)} E_V^3\right)^2$. As the pump fluence increases, the calculated THz third harmonics intensity initially rises and then declines (Fig. 3g). In the increased region with the pump fluence below 4.2 μJ cm$^{-2}$, the increase in third harmonic intensity is primarily driven by the significant rise in $\chi_{loc}^{(3)}$, while the effect of the weakened local field is comparatively minor. In the region where the pump fluence exceeds 4.2 μJ cm$^{-2}$, the growth of $\chi_{loc}^{(3)}$ slows. The cubic dependence between the third harmonic field and the local THz field makes the reduction in $E_V$ a dominant factor, resulting in decreased third-harmonic intensity as the pump fluence continues to increase. The excellent agreement between the measured and calculated third harmonic trends supports our theory of photoinduced modulation of THz THG and the associated analysis of nonlinear carrier dynamics. This insight opens new avenues for the design, optimization, and application of the optically controlled highly efficient THz frequency multipliers and modulators.

Guided by the theoretical model and a detailed analysis of the photo-engineering electron generation and transition processes, the metasurface exhibits photoinduced modulation of nonperturbative carrier distribution under the local strong THz field. This behavior enables an optically controllable nonperturbative THz nonlinear response. A key feature of this response is the optically reconfigurable dependence of the generated third harmonic field strength ($|E_{3f}|$) on the fundamental THz field strength. This dependence was experimentally measured under varying optical pump fluence, as illustrated in Fig. 4a-c. The third-harmonic field strength increases with the fundamental THz field, and the growth rate accelerates as the optical pump fluence rises. Specifically, the metasurface exhibits power law behaviors of $|E_{3f}| \propto |E_0|^l$, where $l$ is 1.6, 1.9, and 2.7 corresponding to pump fluences of 1.5 μJ cm$^{-2}$, 3.5 μJ cm$^{-2}$, and 8.5 μJ cm$^{-2}$, respectively. These dependencies notably deviate from the cubic law, indicating a nonperturbative nonlinear response. Similar nonperturbative nonlinear behavior has previously only been observed in the generation of harmonics excited by high power lasers[44-48]. At low pump fluence, the total carrier density is insufficient to significantly influence the metasurface resonance, resulting in a dramatically enhanced local THz field and a high transition rate. As shown in Fig. 4d, calculations of carrier dynamics reveal that most of the photogenerated carriers in the X-valley at this stage acquire sufficient kinetic energy from the strong local THz field to transition into adjacent satellite valleys. The nonperturbative carrier distribution in the metasurface leads to pronounced nonperturbative nonlinearity, with the dependence of generated third harmonic field strength on the fundamental THz field strength being much smaller than the cubic law. As shown in Fig. 4e, f, increasing the pump fluence weakens the local THz electric field, resulting in lower kinetic energy of the carriers. Consequently, the carrier transition rate between valleys decreases, and the carrier density in the X-valley approaches the total carrier density. The nonperturbative nonlinearity of the metasurface gradually transitions toward perturbative nonlinearity, with the field strength dependence approaching the cubic law (see Note S3 in Supplementary Material). We experimentally confirmed the feasibility of the optically reconfigurable field strength dependence of the third harmonic on the fundamental THz pump and provided a corresponding theoretical model to guide this process. Such photoinduced modulation of nonperturbative THz nonlinear responses has been relatively unexplored in previous studies. It not only offers a convenient platform for exploring the nonperturbative nonlinearity of materials, but positions our metasurface as a strong candidate for nonlinear activation functions in all-optical neural networks. Multiple nonlinear activation functions could be achieved by applying different optical pump fluences to the metasurface. Considering the extremely large FoM of the modulation as demonstrated above, this optically reconfigurable activation function strategy has the potential to significantly enhance training speed, calculation accuracy, and many other key parameters in artificial intelligence [7,49].

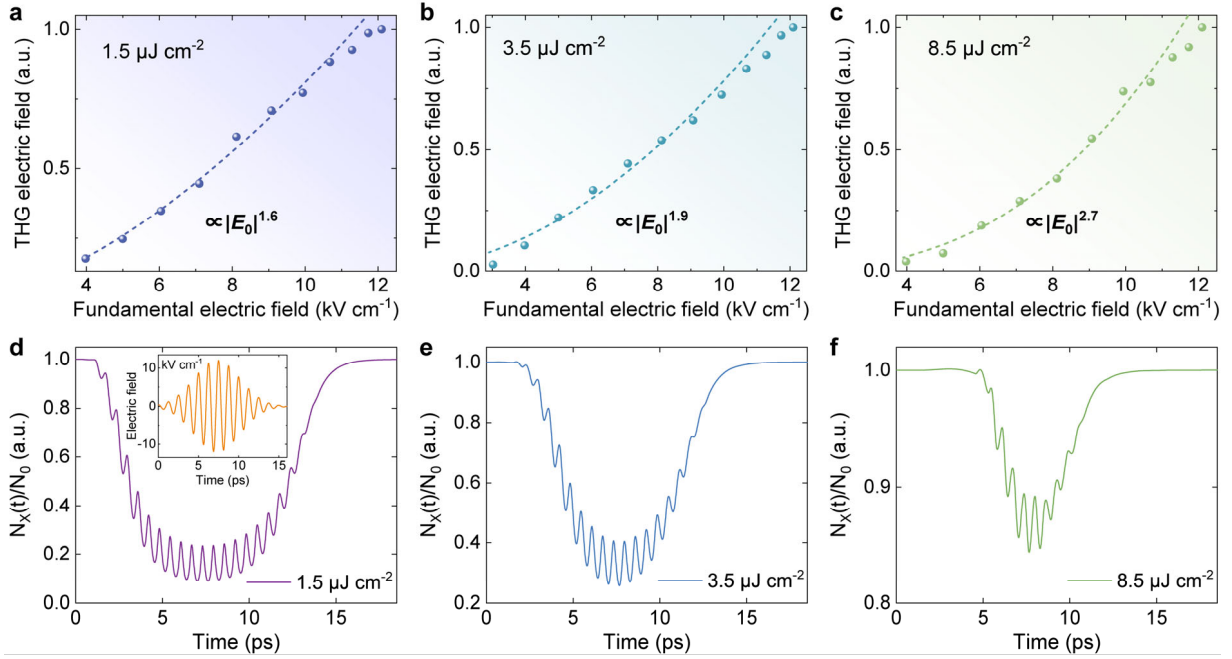

**Fig. 4. Photoinduced modulation of the field dependence of the THz THG on the fundamental pump.** (**a-c**) Dependence of the third harmonic field strength on the fundamental THz field strength. The optical pump fluences are 1.5 μJ cm$^{-2}$ (**a**), 3.5 μJ cm$^{-2}$ (**b**), and 8.5 μJ cm$^{-2}$ (**c**). Dots mark the measured data and the dashed lines mark the theoretical fitting. (**d-f**) Calculated the temporal variations of the normalized carrier density in the X-valley resulting from intervalley transition, with the driving multicycle THz signal shown in the inset of (**d**).

## 3 Discussions

Our metasurfaces demonstrated low-threshold excitation and efficient modulation of THz nonlinearity, with the intensity of the THz third harmonic in the metasurface showing a rapid rise to a peak within a specific range of optical pump fluence, while approaching zero outside this range, resembling a peak function. This behavior suggests potential applications in sharpening and enhancement processing for THz imaging technology. Our results not only demonstrate the modulation of THz nonlinearities but also establish a versatile platform for extending the application of mature optical modulation techniques, such as spatial light modulators, to the THz frequency range. This advancement facilitates the development of sophisticated hybrid multi-physics platforms that can process optical and THz signals simultaneously. Information can be easily encoded into the optical pump pulse using optical modulation technology. Upon receiving the optical pump pulse carrying the original information, the metasurface could efficiently transfer the information to the THz third harmonics. This capability is expected to significantly enhance the efficiency of information transmission between the optical and THz frequency ranges. The SRRs-Si metasurface employed in this study serves as an initial proof-of-concept demonstration. Our approach can be extended to other metasurface structures and semiconductors responsive to external optical, electrical, or thermal stimuli, paving the way for the development of high-performance, tunable, integrated THz nonlinear devices. Furthermore, we believe that the potential of THz nonlinear modulation using metasurfaces remains underexplored. With their high design flexibility, metasurfaces are expected to offer a platform for the polarization and phase modulation of THz harmonics[2,9].

In conclusion, we have demonstrated a novel approach to actively control various THz nonlinear effects through photoinduced modulation of the valley degree of freedom in semiconductor-based metasurfaces. By simultaneously manipulating the presence of free carriers, the population of carriers across different energy valleys, and the local THz field strength using photonic methods, several THz nonlinear effects, including nonlinear transmission and THG, can be efficiently enhanced and modulated within a single metasurface at room temperature. This approach achieves an ultra-high modulation depth exceeding $2\times10^4$% under very weak optical energy at the level of picojoules. Furthermore, we demonstrate optical reconfigurations of fundamental pump dependence of THG by manipulating the nonperturbative THz nonlinear behavior of free electrons, indicating a promising candidate for nonlinear activation functions in all-optical neural networks. Our findings provide a robust foundation for dynamic nonlinear devices with the potential of on-chip integration, such as highly integrated THz switches, frequency converters, modulators, and sources, paving the way for advancements in high-speed THz communications and all-optical computing.

## 4 Methods

### 4.1 Sample fabrication

The designed metasurface structure was fabricated on a commercially available SoS wafer composed of a 500 nm thick epitaxial (001) Si layer formed on a 460 μm thick R-plane sapphire substrate. The resistivity of the Si layer was 100 Ω·cm according to the manufacturer. The fabrication of the metasurface sample began with the patterning of metallic patterns on the Si layer. First, a layer of photoresist was spin-coated on the substrate to form the SRR structures through a standard ultraviolet photolithography process. Next, a 500 nm thick aluminum film was deposited via electron beam evaporation. The last step was the lift-off of the residual photoresist in acetone, leading to the final samples.

## 4.2 Experimental Setup and Measurement

The room-temperature optically tunable transmission spectra were measured using table-top THz time-domain spectroscopy based on a femtosecond laser pumping nonlinear organic crystal for generation and a GaP crystal for detection. The samples were collinearly illuminated with the incident THz pulse and a pump laser pulse. The time delay was fixed on 10 ps between the pump laser beam and the incident THz wave. The broadband frequency-dependent transmission spectrum is calculated using normalized transmission, $T = |E_{ms}(f)/E_0(f)|$, where $E_{ms}(f)$ and $E_0(f)$ are Fourier-transformed frequency domain electric field spectra of the metasurface and the air, respectively.

The room-temperature optical switching and modulation of THG were measured using the above THz source. To enhance measurement accuracy, two 0.8 THz bandpass filters were employed, effectively suppressing the harmonic background of the THz source and providing a maximum vertical field strength of 19.2 kV cm$^{-1}$ with a central frequency of 0.8 THz. A set of wire-grid polarizers were used to modulate the fundamental THz pump. Similar to the transmission spectra measurements, the samples were collinearly illuminated with both the incident THz pulse and a pump laser pulse, with a fixed time delay of 10 ps to mitigate unintended frequency conversion from the rapidly time-varying metamaterial. The transmitted wave of the sample passing through a 2.4 THz bandpass filter was measured using EOS with a 1 mm thick zinc telluride (ZnTe) crystal. Throughout all experiments, the relative humidity was maintained below 5% through dry N₂ purging (see Note S2 in Supplementary Material for details).

### 4.3 Numerical simulation

The metasurfaces were numerically modeled and simulated using a commercial finite-element package (COMSOL Multiphysics). Aluminum was modeled as a lossy metal with a conductivity of $3.72\times10^7$ S m$^{-1}$. The Si film beneath the SRRs was modeled as a dielectric with $\varepsilon_{Si}$=11.7, and the conductivity of the Si was varied to simulate the effect of optical excitation (see Note S1 in Supplementary Material for details).

### Data availability

The data supporting the findings of this study are available from the corresponding author upon reasonable request.

### Declaration of competing interest

The authors declare that they have no known competing financial interests or personal relationships that could have appeared to influence the work reported in this paper.

### Acknowledgements

This work was supported by the National Key R&D Program of China (2023YFB3811400), Basic Science Center Project of NSFC (52388201), National Natural Science Foundation of China (52332006, 12504387, and 52573329), Beijing Natural Science Foundation (Z240008), and the China Postdoctoral Science Foundation (BX20250299 and 2025M773396).